%% file: cmd3_etaprim.tex
\documentclass[12pt]{elsarticle}
\usepackage{refmerge}
\usepackage{epsfig}
\usepackage{psfrag}
\usepackage{amsmath}
\def\Ecm {\ensuremath{\rm E_{\rm c.m.}}\ }
\def\BR {\ensuremath{\mathcal B}\ }
\def\epem {\ensuremath{e^+ e^-}\ }
\def\pipi {\ensuremath{\pi^+\pi^-}\ }

\def\piz {\ensuremath{\pi^0}\ }
\begin{document}
%\begin{frontmatter}
\date{\today}

\title{\bf{ \boldmath
SEARCH FOR THE PROCESS $e^+e^-\to \eta^\prime(958)$ WITH THE CMD-3 DETECTOR
}}

\input{authors_cmd3.tex}

%========================================================================

%
\vspace{0.7cm}
%\newpage
\begin{abstract}
\hspace*{\parindent}
A search for the process $\epem\to \eta^\prime(958)$ in the 
$\pipi\eta\to\pipi\gamma\gamma$  final state has been performed with 
the CMD-3 detector at the VEPP-2000 $e^+e^-$ collider. Using an integrated 
luminosity of  2.69 pb$^{-1}$ collected at the center-of-mass energy 
\Ecm = 957.68 MeV we set an upper limit for the product of electronic width 
and branching fractions 
$\Gamma_{\eta^\prime(958)\to\epem}\cdot\BR_{\eta^\prime(958)\to\pipi\eta}\cdot\BR_{\eta\to\gamma\gamma}<0.00041$ 
eV at 90\% C.L. 
\end{abstract}

%\end{frontmatter}
\maketitle
\baselineskip=17pt
\section{ \boldmath Introduction}
\hspace*{\parindent}
Direct production of C-even resonances in \epem collisions is possible via 
a two-photon intermediate state. A search for direct production of the 
$\eta^\prime(958),~f_0(980),~a_0(980),~f_2(1270),~f_0(1300)$, and $a_2(1320)$ 
was performed with the ND detector at the VEPP-2M collider~\cite{ND}. 
Only upper limits have been set, in particular,
for the electronic width of the $\eta^\prime(958)$ a limit 
$\Gamma_{\eta^\prime(958)\to\epem}<$0.06 eV at 90\% C.L. has been obtained. 
In the unitarity limit, when both photons are assumed to be real, the 
branching fraction of the decay of the $\eta^\prime(958)$, 
denoted below as $\eta^\prime$, 
to an \epem pair, $\BR_{\eta^\prime\to\epem}$,
can be estimated using the two-photon branching fraction 
$\BR_{\eta^\prime\to\gamma\gamma} = 0.0220\pm 0.0008 $ ~\cite{PDG} 
and the expression from Refs.~\cite{Drell, Land}
\begin{equation}
\BR_{\eta '\to\epem} = \BR_{\eta '\to\gamma\gamma}\frac{\alpha^2}{2\beta}
(\frac{m_e}{m_{\eta '}})^2[ln(\frac{1+\beta}{1-\beta})]^2 = (3.75\pm0.14)\times 10^{-11},
\label{bee}
\end{equation}
where $\alpha$ is the fine structure constant, $m_e$ and $m_{\eta^\prime}$
are masses of electron and $\eta^\prime$, respectively,
and $\beta =\sqrt{1-4(\frac{m_e}{m_{\eta^\prime}})^2}$. 
Using  a total width of the $\eta^\prime$, 
$\Gamma_{\eta^\prime} = 0.198\pm 0.009$ MeV~\cite{PDG}, we obtain in the 
unitarity limit  $\Gamma_{\eta^\prime\to\epem} = (7.43\pm0.29)\times 10^{-6}$ eV. 
Photon virtuality  and the $\eta^\prime \to\gamma\gamma$ transition 
form factor can significantly enhance, by a factor of 5-10, the electronic width value, 
as discussed in Ref.~\cite{Land}. 

An observation of the direct production of C-even resonances in 
\epem collisions, and, in particular, of the reaction 
$\epem\to \eta^\prime(958)$, could help to develop theoretical
approaches to a photon-loop calculation, which is a crucial 
point in the estimation of the hadronic light-by-light 
contribution~\cite{lbl1,lbl2} to 
the muon anomalous magnetic moment (g-2)~\cite{g-2}. 

In this paper we report a search for the process  $\epem\to \eta^\prime(958)$ 
in the $\eta^\prime(958)\to\pipi\eta\to\pipi\gamma\gamma$ 
decay chain. The search is based on the 
2.69 pb$^{-1}$ of an integrated luminosity collected with the CMD-3 detector
at the center-of-mass (c.m.) energy of the  VEPP-2000
collider~\cite{vepp1,vepp2} 
close to the nominal $\eta^\prime(958)$ mass: 
$m_{\eta^\prime}=957.78\pm0.06 $ MeV/$c^2$~\cite{PDG}. 
\section{Detector and data taking conditions}
\label{detector}
\begin{figure}[tbh]
\begin{center}
%\vspace{-0.5cm}
\includegraphics[width=1.0\textwidth]{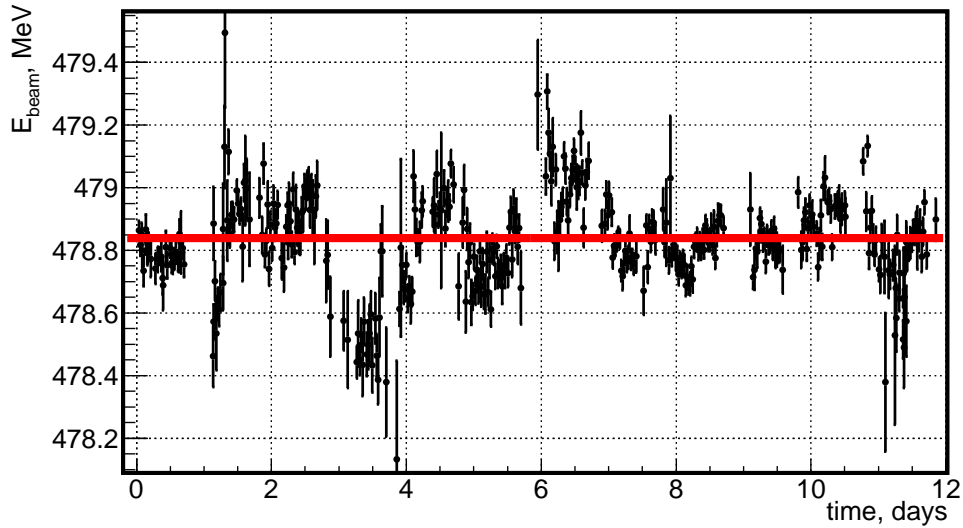}
\vspace{-0.7cm}
\caption
{Beam energy measurements during data taking.
}
\label{ebeam}
\end{center}
\end{figure}

\hspace*{\parindent}
The total width of the $\eta^\prime$ is relatively small, and it is
very important to have  c.m. energy close to this value.
The collider beam energy was continuously monitored during the whole period
of data taking (12 days) using the Back-Scattering-Laser-Light
system~\cite{laser}. 
Figure~\ref{ebeam} shows measurements of the beam energy, $\rm
E_{beam}$, which demonstrate relatively good stability of the
collider energy.  The average value of the  c.m. energy is 
${\rm E}_{\rm c.m.}^{\rm av.}$ = 957.678$\pm$0.014 MeV with a few deviations of
up to 0.2 MeV, corresponding to less than 5\% of the integrated
luminosity,  which are still within an energy spread of the collider
beams as shown below.   

\begin{figure}[tbh]
\begin{center}
\vspace{-0.5cm}
\includegraphics[width=1.0\textwidth]{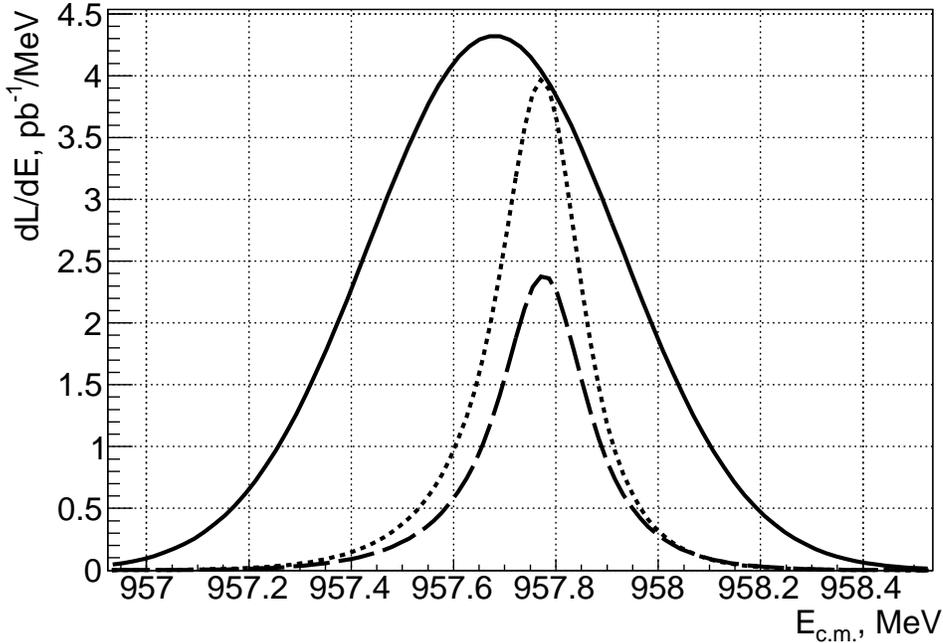}
\vspace{-0.7cm}
\caption
{The luminosity distribution versus the c.m. energy (solid line)
with an area normalized to the 2.69 pb$^{-1}$ of integrated luminosity. The areas
below the dotted and dashed lines illustrate an  ``effective''
luminosity for the $\eta^\prime$ BW  and BW convolved with the 
radiator function, respectively.
}
\label{lumwidth}
\end{center}
\end{figure}
The beams of the collider have an energy spread mainly due to the 
quantum effects of the synchrotron radiation. The c.m. energy spread
of the VEPP-2000 collider $\sigma_{{\rm E}_{\rm c.m.}} = 0.246 \pm 0.030 $~MeV is 
calculated according to Ref.~\cite{Koop}
using a longitudinal distribution of the interaction region 
$\sigma_{\rm Z} = 2.3 $ cm and RF cavity voltage ${\rm V}_{\rm cav} = 18 $ kV. 
Full energy spread (FWHM = 0.590 MeV)
is significantly larger than the width of $\eta^\prime$, as
demonstrated in Fig.~\ref{lumwidth}. A solid line in
Fig.~\ref{lumwidth} shows a differential luminosity distribution
${\rm dL}/{\rm dE}$ versus the c.m. collision energy, with an area 
normalized to the total integrated luminosity L = 2.69 pb$^{-1}$.   
This distribution should 
be convolved with a Breit-Wigner (BW) function, describing 
the $\eta^\prime$ line shape (the dotted line in Fig.~\ref{lumwidth}).
Radiation of photons by initial particles changes a collision
energy and the BW should be additionally convolved with a radiator
function described in Ref.~\cite{kur_fad,rmc}. These radiative
corrections decrease the number of signal events by approximately 40\%.
The area under the dashed line in Fig.~\ref{lumwidth} illustrates 
the ``effective'' integrated luminosity for $\eta^\prime$
production in our experiment in comparison with the total integrated 
luminosity under the solid line.

If we describe the $\eta^\prime$ production cross section as
\begin{equation}
\sigma^f(E) = \frac{\sigma^0_{\eta^\prime\to f} m^2_{\eta^\prime}\Gamma_{\eta^\prime}^2}
{(m^2_{\eta^\prime}-E^2)^2+E^2\Gamma^2_{\eta^\prime}},
\label{sigma0bw}
\end{equation}
where $\sigma^0_{\eta^\prime\to f}$ is the peak cross section for the process
$\epem\to\eta^\prime$ with $\eta^\prime$ decay to the final 
state $f$, we can calculate an integrated production cross section as
\begin{equation} 
\sigma_{\rm int}^f = \int_{0}^{E_{\rm beam}}dE\int_{0}^{1}\frac{1}
{\sqrt{2\pi}\sigma_{\Ecm}}e^{-\frac{( \Ecm^{\rm av.}-E)^2}{2\sigma_{\Ecm}^2}}\cdot F(x,E)\cdot\sigma^f(E(1-x))dx~,
\label{sigmavis}
\end{equation}
where $F(x,E)$ is the radiator function~\cite{kur_fad,rmc}, and $x$ is a 
fraction of energy taken by photons. 
Using a relation
\begin{equation}
\sigma^0_{\eta^\prime\to f}\Gamma_{\eta^\prime} = 
4\pi\frac{{\rm C}\cdot\BR_{\eta^\prime\to f}\cdot\Gamma_{\eta^\prime\to\epem}}
{m^2_{\eta^\prime}},
\label{sigma0}
\end{equation}
where   $\BR_{\eta^\prime\to f}$ is the branching fraction to the measured 
final state $f$, and ${\rm C}=3.89\cdot 10^{11}~\rm {nb~MeV^2}$ is a
conversion constant~\cite{PDG}, we perform the integration of 
Eq.~\ref{sigmavis}, and obtain 
$\sigma^f_{\rm int} =(6.38\pm0.23)\cdot\Gamma_{\eta^\prime\to\epem}({\rm
  eV})\cdot\BR_{\eta^\prime\to f}$ nb for our experimental
conditions. The error in the coefficient reflects uncertainty 
in the $\sigma_{{\rm E}_{\rm c.m.}}$ value due to variation of cavity voltage ${\rm V}_{\rm cav}$,
and uncertainty in the beam energy ($\Ecm^{\rm av.}$) measurements, 
including energy instability according to Fig.~\ref{ebeam},
weighted with the fraction of integrated luminosity during the energy shifts.
Since the energy spread is significantly larger than 
the total $\eta^\prime$ width, the integrated cross section is
proportional to the product of the $\eta^\prime$ electronic width and 
branching fraction to the measured final state $f$, calculated as
\begin{equation}
\Gamma_{\eta^\prime\to\epem}\cdot\BR_{\eta^\prime\to f}=\frac{\rm N}{6.38\cdot\epsilon^f\cdot {\rm L}}~({\rm eV}),
\label{result}
\end{equation}
where  $\rm N$ is the number of observed signal events and $\epsilon^f$ is 
a detection efficiency for the final state $f$.

For the studied decay mode with $\BR_{\eta^\prime\to\pipi\eta} = 0.429$~\cite{PDG}, 
the calculated cross section for the unitarity limit of 
$\Gamma_{\eta^\prime\to\epem}$  is very small,  
$\sigma_{\rm int}\approx 2\times 10^{-5}$~nb, and should be compared with the
cross section of the single-photon reaction $\epem\to\pipi\eta$. The events of 
this process were observed only above $\Ecm > 1200$ MeV and the 
energy dependence of the cross section was well described
in the model where the final state is produced via three interfering
resonances -- $\rho(770),~\rho(1450)$ and $\rho(1700)$, see for example Ref.~\cite{simon}.
%by the $\epem\to\rho(770,1450,1600)\to\pipi\rho(770)$ model~\cite{simon}. 
At $\Ecm = 958$ MeV we obtain 
$\sigma_{\epem\to\pipi\eta} = (1.2\pm0.6)\times 10^{-3} $ nb
with large uncertainty from the unknown interference phases between the
$\rho$ resonances extrapolating to the region close to the threshold. This value 
is comparable or larger than the two-photon cross section $\sigma_{\rm int}$ 
with possible enhancement due to the form factor,
 and to prove signal observation a measurement outside 
$\eta^{\prime}(958)$ mass is needed.
Note that the $\eta^\prime\to\pi^0\pi^0\eta$ decay mode is free from the 
single-photon physical background, 
and we plan to use it for such a study as well.

The general-purpose detector CMD-3 has been described in 
detail elsewhere~\cite{sndcmd3}. Its tracking system consists of a 
cylindrical drift chamber (DC)~\cite{dc} and double-layer multiwire 
proportional Z-chamber, both also used for a trigger, and both inside a thin 
(0.2~X$_0$) superconducting solenoid with a field of 1.3~T.
The liquid xenon (LXe) barrel calorimeter with 5.4~X$_0$ thickness has
fine electrode structure, providing 1-2 mm spatial resolution~\cite{lxe}, and
shares the cryostat vacuum volume with the superconducting solenoid.     
The barrel CsI crystal calorimeter~\cite{lxe} with a thickness 
of 8.1~X$_0$ is placed outside  the LXe calorimeter,  and the end-cap 
BGO calorimeter with a thickness of 13.4~X$_0$ is placed inside the 
solenoid~\cite{cal}.
The luminosity is measured using events of Bhabha scattering 
at large angles~\cite{lum}. 
\section{Selection of $\epem\to\pipi\gamma\gamma$ events}
\label{select}
\hspace*{\parindent}
Candidates for the process under study are required to have 
two good charged-particle tracks and two or more  clusters in the calorimeters
not related to tracks assumed to be photons. We use the following 
``good'' track definition:
\begin{itemize}
\item{
A track contains more than five hits in the DC.
}
\item{
A track momentum is larger than 40~MeV/$c$.
}
\item{
A minimum distance from a track to the beam axis in the
transverse  plane is less than 0.5 cm.
}
\item{
A minimum distance from a track to the center of the interaction region along
the beam axis Z  is less than 10 cm.
}
\item{
A track has a polar angle large enough to cross half of the DC radius.
}
\end{itemize} 
\begin{figure}[tbh]
\begin{center}
\vspace{-0.5cm}
\includegraphics[width=1.0\textwidth]{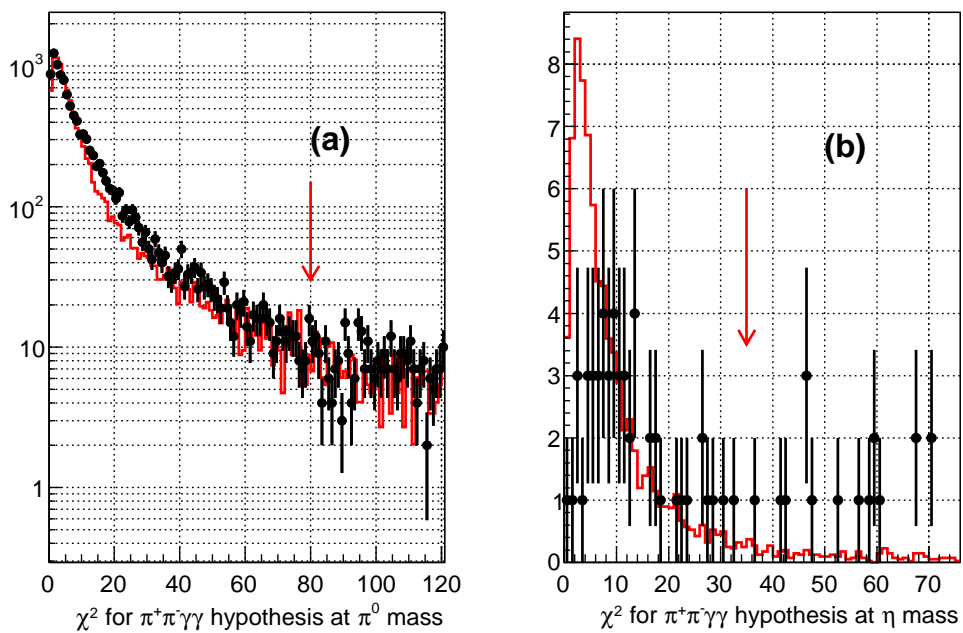}
\vspace{-0.7cm}
\caption
{
The $\chi^2$ distribution of events with two tracks and two photons for the
$\epem\to\pipi\gamma\gamma$ hypothesis for data (dots) and 
corresponding simulations (histograms), when the two-photon invariant
mass is in  the $\pm$35~MeV/$c^2$ window around the $\piz$ mass (a) and
in the $\pm$35~MeV/$c^2$ window around the $\eta$ mass (b). 
}
\label{chi2}
\end{center}
\end{figure}
Two tracks are required to have opposite charges and are assumed to be
pions. The detected photons are required to have more than 25 MeV
energy deposition in the calorimeters.   Reconstructed momenta and
angles of the detected charged tracks and energy and angles of two
photons are subject to the kinematic fit, assuming that the total
energy is equal to \Ecm and total momentum is equal to zero. 
The covariance matrices for charged tracks and photons are used in the fit 
and provide a $\chi^2$ value for each event. If an event candidate has 
more than two photons, the photon pair with the smallest $\chi^2$ value 
is retained. As a result of the fit, we obtain improved values of the 
momenta, energies and angles for all particles. 
The main contribution to the selected sample comes from the
process $\epem\to\pipi\piz\to \pipi\gamma\gamma$. We perform simulation of the  
processes $\epem\to\pipi\piz$ and $\epem\to\eta^\prime$ and apply all 
experimental conditions and selections to the simulated samples. 
We use the $\epem\to\pipi\piz$ events to verify our simulation.       
Figure~\ref{chi2}(a) shows the $\chi^2$ distributions for the
experimental (dots) and simulated $\epem\to\pipi\piz$ (histogram)
events when the invariant mass of the photon pair is in the $\pm$35~MeV/$c^2$
window around the $\piz$ mass. A vertical arrow shows the applied
selection.  Figure~\ref{chi2}(b) presents the $\chi^2$ 
distributions for the event candidates, in which the invariant mass of 
the photon pair is in the $\pm$35~MeV/$c^2$ window around the $\eta$ 
mass for data (points), and the simulated distribution for the process
$\epem\to\eta^\prime \to\pipi\eta\to \pipi\gamma\gamma$ 
(histogram). 
The  $\chi^2$ distributions for data and simulation for the $\piz$ signal in Fig.~\ref{chi2}(a)
are in good agreement, and tighter selection shown
by the vertical arrow in Fig.~\ref{chi2}(b) changes the number of $\pipi\piz$ signal events by 1\% only. After this selection some 
candidate events for the process $\epem\to\eta^\prime$ are observed.
\begin{figure}[tbh]
\begin{center}
\vspace{-0.5cm}
\includegraphics[width=1.0\textwidth]{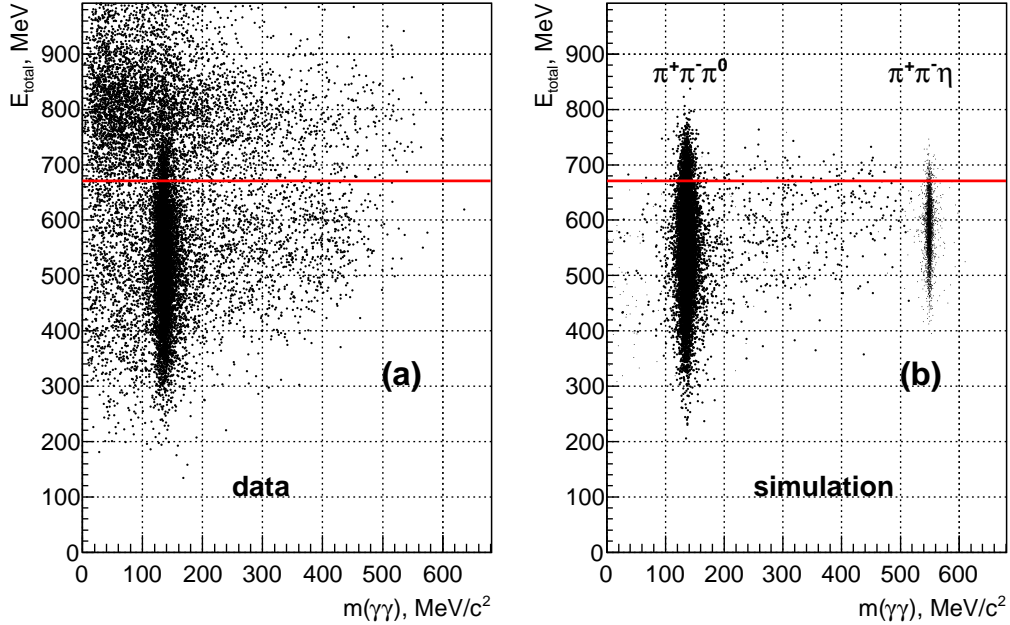}
\vspace{-0.7cm}
\caption
{
Scatter plot of the total energy deposition versus two-photon invariant mass 
for selected experimental $\pipi\gamma\gamma$ events (a) and simulated  
$\pipi\piz\to\pipi\gamma\gamma$ and $\eta^\prime\to\pipi\eta\to\pipi\gamma\gamma$ 
events (b). Solid lines show applied selection.
}
\label{mvsen}
\end{center}
\end{figure}

Figure~\ref{mvsen}(a) shows a scatter plot of the total energy 
deposition in the calorimeter, ${\rm E}_{\rm total}$, versus two-photon
invariant mass for selected experimental $\pipi\gamma\gamma$ events. 
A clear signal of the $\pipi\piz$ events is seen. Simulated events for 
the processes $\epem\to\pipi\piz\to\pipi\gamma\gamma$ and 
$\epem\to\eta^\prime\to\pipi\eta\to\pipi\gamma\gamma$ are shown in 
Fig.~\ref{mvsen}(b). The experimental sample contains a relatively
large  number of background events from various quantum electrodynamics 
processes, when scattered electron and positron are accompanied with a 
number of radiative photons. These processes leave significant amount of energy 
in the calorimeter and can be effectively suppressed by requiring 
${\rm E}_{\rm total} < 0.7\cdot\Ecm$  (solid lines in Fig.~\ref{mvsen}), retaining 
87\% of signal events in good agreement with simulation.

We calculate the detection efficiency from the 
$\epem\to\pipi\eta\to\pipi\gamma\gamma$ simulated events
as a ratio of the number of events after selections described above 
to the total number of generated events and obtain 
$\epsilon^f = 31.1$\% for this final state.  

 Using the signal from the process $\epem\to\pipi\piz\to\pipi\gamma\gamma$,
and taking into account 1\% data-simulation corrections for charged tracks and 2\% 
for the photons, 
we found that overall data-simulation discrepancy 
gives less than 5\% uncertainty in the resulting efficiency for applied
selections. 

\begin{figure}[tbh]
%\begin{center}
%\vspace{-0.5cm}
\includegraphics[width=1.0\textwidth]{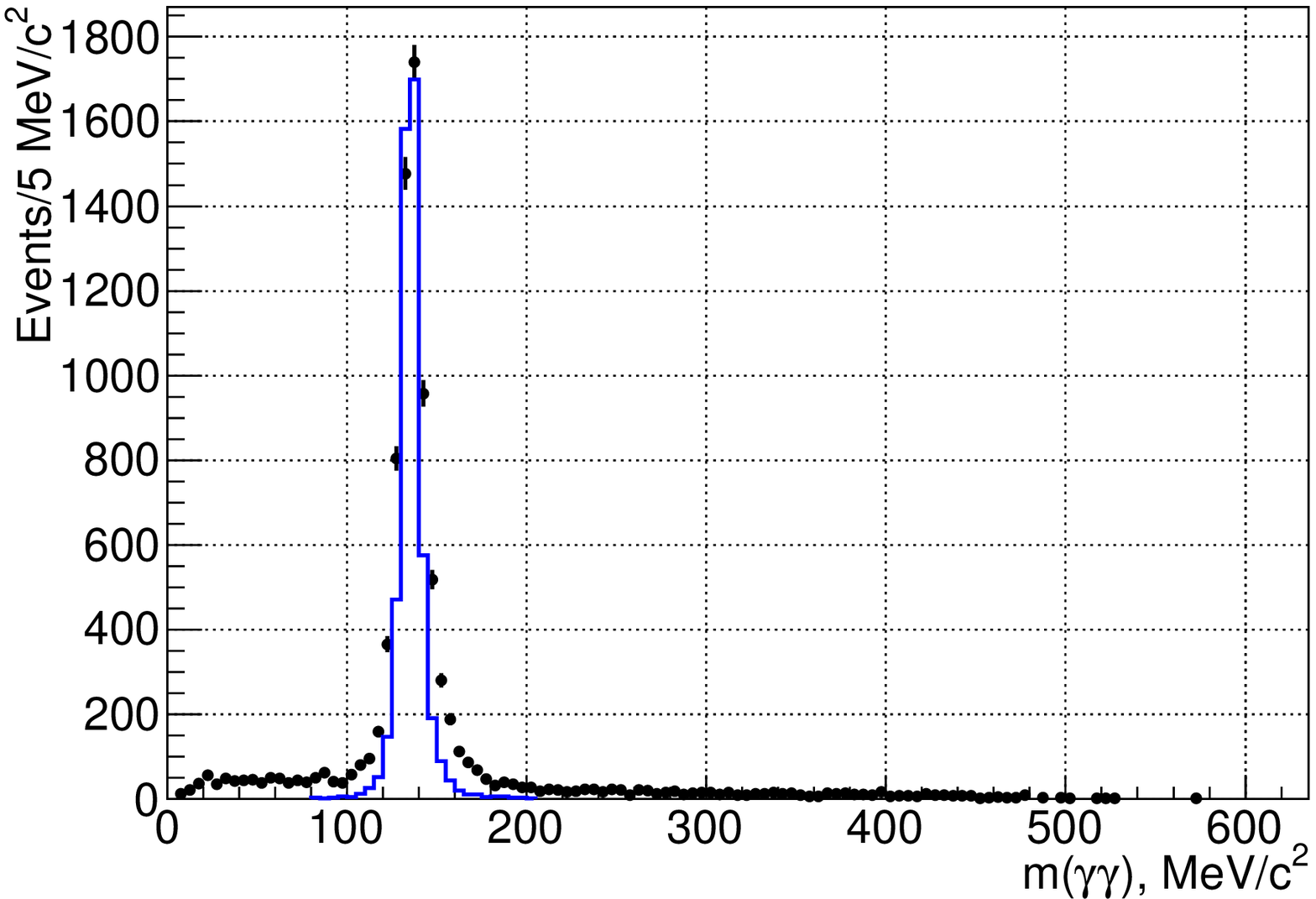}
\hspace{-8.8cm}
\includegraphics[width=0.5\textwidth]{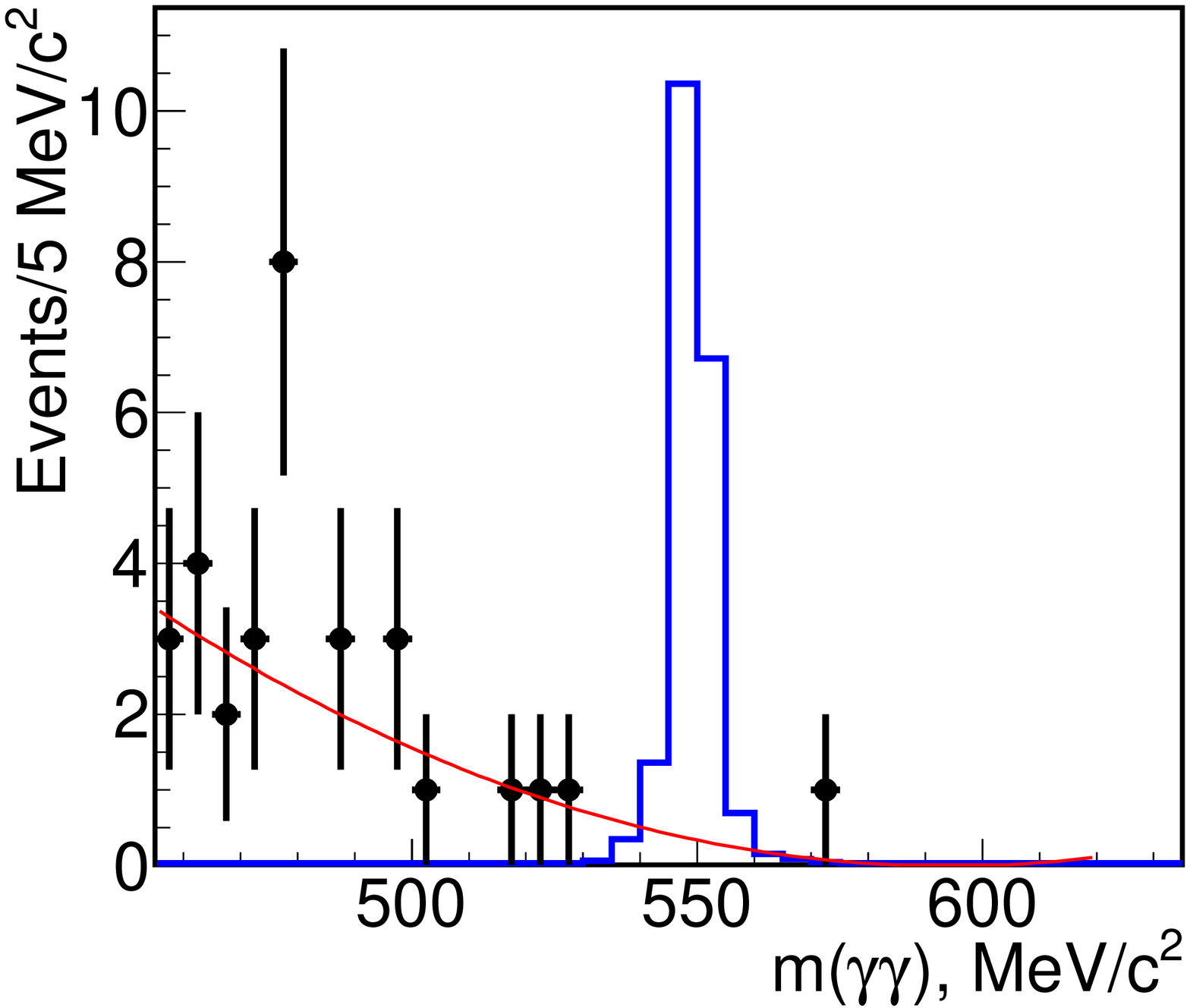}
\vspace{-0.3cm}
\caption
{
Projection plot of Fig.~\ref{mvsen} after the applied cut on total
energy. Dots are from data, the histogram shows invariant mass of 
two photons from the simulated process $\epem\to\pipi\piz\to\pipi\gamma\gamma$. 
The expanded view of the 450-650~MeV/$c^2$ region is shown in
the box with the signal shape of the process
$\epem\to\eta^\prime\to\pipi\eta\to\pipi\gamma\gamma$ 
(histogram)  expected from simulation,
and expected level of background (line).  
}
\label{mgg}
%\end{center}
\end{figure}

Figure~\ref{mgg} shows a projection plot of Fig.~\ref{mvsen}(a) after 
the applied selections. Dots are for data, and the histogram shows the 
invariant mass of two photons from the simulated process
$\epem\to\pipi\piz\to\pipi\gamma\gamma$. The expanded view 
of the 450-650~MeV/$c^2$ region is shown in the box with 
the shape of signal for the process $\epem\to\eta '\to\pipi\eta\to\pipi\gamma\gamma$
(histogram) obtained from the simulation. The line shows the 
expected level of background, obtained from the fit of events in this 
region with a second-order polynomial function. 
We found no event candidates for this process in the signal region, 
and estimate the expected background of $1.0\pm0.5$ events.
We use a conservative estimate of
the number of signal events as N$<$2.0 at
90\% C.L. using the Feldman-Cousins approach~\cite{felcou}, assuming no observed events and
expected background of 0.5 events.

Using Eq.~\ref{result} we obtain the result
$$
\Gamma_{\eta^\prime\to e^+e^-}\BR_{\eta^\prime\to\pipi\eta}
\BR_{\eta\to\gamma\gamma}<\frac{2.0}{6.38\cdot 0.311\cdot 2690} = 0.00037~{\rm eV~at~90\%~C.L.}.
$$

After taking into account the systematic uncertainties on luminosity (2\%),
efficiency (5\%), and beam energy instability (5\%) we conservatively increase
the upper limit by 12\%:
$$
\Gamma_{\eta^\prime\to e^+e^-}\BR_{\eta^\prime\to\pipi\eta}
\BR_{\eta\to\gamma\gamma}< 0.00041~{\rm eV~at~90\%~C.L.}.
$$
The decay rates of $\eta^\prime\to\pipi\eta$ and
$\eta\to\gamma\gamma$ are relatively well known, and using the values 
from Ref.~\cite{PDG} we obtain $\Gamma_{\eta^\prime\to e^+e^-}<0.0024$~eV 
at 90\% C.L..
Finally, from the total width of the $\eta^\prime$ from Ref.~\cite{PDG} 
we calculate $\BR_{\eta^\prime\to e^+e^-} < 1.2\times 10^{-8}$,
that should be compared to the value from the ND experiment~\cite{ND}, 
$\BR_{\eta^\prime\to e^+e^-} < 2.1\times 10^{-7}$ listed in the PDG 
tables~\cite{PDG}. The latter value was obtained from the limit 
$\Gamma_{\eta^\prime\to e^+e^-}<$ 0.06~eV and $\eta^\prime$ width of about
300~keV known at that time. 

In our experiment we can also set a limit on the cross section for the process 
$\epem\to\pipi\eta$ at \Ecm = 957.7 MeV, which is found to be 
$\sigma(\epem\to\pipi\eta) = N/(\epsilon_{\pipi\eta}L) < 6.1$ pb for 90 C.L., 
where $\epsilon_{\pipi\eta} = \epsilon^f\cdot \BR_{\eta\to\gamma\gamma} = 0.122$ 
is a detection efficiency for this process. 

\section*{ \boldmath Conclusion}
\hspace*{\parindent}
We search for direct production of the C-even $\eta^\prime(958)$ meson in 
\epem collisions. A special experimental run of the VEPP-2000
collider was performed at the c.m. energy close to the $\eta^\prime$
mass with a 2.69 pb$^{-1}$ integrated luminosity.
We found no event candidates for the process 
$\epem\to\eta^\prime(958)\to\pipi\eta\to \pipi\gamma\gamma$ 
and obtain the upper limit for the product 
$\Gamma_{\eta^\prime(958)\to e^+e^-}\BR_{\eta^\prime\to\pipi\eta}\BR_{\eta\to\gamma\gamma}<0.00041$ eV. 
This limit is ten times lower compared to the previous measurement~\cite{ND}.
\subsection*{Acknowledgements}
\hspace*{\parindent}
The authors are grateful to A. Kup{\'s}{\'c} for stimulating discussions 
and to M.\ N.\ Achasov for help with energy calibration.
We thank the VEPP-2000 personnel for the excellent machine operation. 

This work is supported in part by the Russian Education and Science
Ministry (grant N 14.610.21.0002, identification number RFMEFI61014X0002), 
by the Russian Foundation for Basic Research grants   
RFBR 13-02-00991-a, 
RFBR 13-02-00215-a,
RFBR 12-02-01032-a, 
RFBR 13-02-01134-a,
RFBR 14-02-00580-a, 
RFBR 14-02-31275-mol-a,
RFBR 14-02-00047-a,
RFBR 14-02-31478-mol-a,
RFBR 14-02-91332 
and the DFG grant HA 1457/9-1.

\input{biblio_etaprim_cmd3.tex}
\end{document}

%% file: authors_cmd3.tex
\author[adr1,adr2]{R.R.Akhmetshin}
\author[adr1,adr2]{A.V.Anisenkov}
\author[adr1,adr2]{V.M.Aulchenko}
\author[adr1]{V.Sh.Banzarov}
%\author[adr1]{L.M.Barkov}
\author[adr1]{N.S.Bashtovoy}
\author[adr1,adr2]{D.E.Berkaev}
\author[adr1,adr2]{A.E.Bondar}
\author[adr1]{A.V.Bragin}
\author[adr1,adr2]{S.I.Eidelman}
\author[adr1,adr6]{D.A.Epifanov}
\author[adr1,adr3]{L.B.Epshteyn}
\author[adr1]{A.L.Erofeev}
\author[adr1,adr2]{G.V.Fedotovich}
\author[adr1,adr2]{S.E.Gayazov}
\author[adr1,adr2]{A.A.Grebenuk}
\author[adr1,adr2,adr3]{D.N.Grigoriev}
\author[adr1]{E.N.Gromov}
\author[adr1,adr2]{F.V.Ignatov}
\author[adr1]{S.V.Karpov}
\author[adr1,adr2]{V.F.Kazanin}
\author[adr1,adr2]{\fbox{B.I.Khazin}}
%\author[adr1]{A.N.Kirpotin}
\author[adr1,adr2]{I.A.Koop}
\author[adr1,adr2]{O.A.Kovalenko}
\author[adr1]{A.N.Kozyrev}
\author[adr1,adr2]{E.A.Kozyrev}
\author[adr1,adr2]{P.P.Krokovny}
\author[adr1,adr2]{A.E.Kuzmenko}
\author[adr1,adr2]{A.S.Kuzmin}
\author[adr1,adr2]{I.B.Logashenko}
\author[adr1,adr2]{P.A.Lukin}
%\author[adr1]{A.P.Lysenko}
\author[adr1]{K.Yu.Mikhailov}
\author[adr1,adr2]{N.Yu.Muchnoi}
\author[adr1]{V.S.Okhapkin}
\author[adr1]{Yu.N.Pestov}
\author[adr1,adr2]{E.A.Perevedentsev}
\author[adr1,adr2]{A.S.Popov}
\author[adr1,adr2]{G.P.Razuvaev}
\author[adr1]{Yu.A.Rogovsky}
\author[adr1]{A.L.Romanov}
\author[adr1]{A.A.Ruban}
\author[adr1]{N.M.Ryskulov}
\author[adr1,adr2]{A.E.Ryzhenenkov}
\author[adr1,adr2]{V.E.Shebalin}
\author[adr1,adr2]{D.N.Shemyakin}
\author[adr1,adr2]{B.A.Shwartz}
\author[adr1,adr2]{D.B.Shwartz}
\author[adr1,adr4]{A.L.Sibidanov}
\author[adr1]{P.Yu.Shatunov}
\author[adr1]{Yu.M.Shatunov}
%\author[adr1]{\fbox{I.G.Snopkov}}
\author[adr1,adr2]{E.P.Solodov\fnref{tnot}}
\author[adr1]{V.M.Titov}
\author[adr1,adr2]{A.A.Talyshev}
\author[adr1]{A.I.Vorobiov}
\author[adr1,adr2]{Yu.V.Yudin}
%\author[adr1,adr5]{A.C.Zaytsev}
%\author[adr1]{Yu.M.Zharinov}

\address[adr1]{Budker Institute of Nuclear Physics, SB RAS, 
Novosibirsk, 630090, Russia}
\address[adr2]{Novosibirsk State University, Novosibirsk, 630090, Russia}
\address[adr3]{Novosibirsk State Technical University, 
Novosibirsk, 630092, Russia}
\address[adr4]{University of Sydney, School of Physics, 
Falkiner High Energy Physics, NSW 2006, Sydney, Australia}
%\address[adr5]{Brookhaven National Laboratory, P.O. Box 5000 Upton, 
%NY 11973-5000, USA}
\address[adr6]{University of Tokyo, Department of Physics, 
7-3-1 Hongo Bunkyo-ku Tokyo, 113-0033, Japan}
\fntext[tnot]{Corresponding author: solodov@inp.nsk.su}